\documentclass[12pt]{article}
\usepackage{jheppub}

\def\CN {{\cal N}}
\def\CO {{\cal O}}

\def\CO {{\cal O}}

\title{Boundary F-maximization}

\author[1]{Davide Gaiotto}

\affiliation[1]{Perimeter Institute for Theoretical Physics\\%
31 Caroline Street North, ON N2L 2Y5, Canada}

\abstract{We discuss a variant of the F-theorem and F-maximization principles which applies to (super)conformal boundary conditions of 4d (S)CFTs.}

\begin{document}
\maketitle

\section{Introduction}

The RG flows between three-dimensional conformal field theories are constrained by an F-theorem, conjectured by \cite{Myers:2010xs,Myers:2010tj,Jafferis:2011zi} and proven by \cite{Casini:2012ei}. 
The quantity $F$ which should decrease along the RG flow is defined either by the finite part of the free energy on $S^3$ or by the finite part of the entanglement entropy for a disk and its complement
 \cite{Casini:2011kv}.

In theories with $\CN=2$ supersymmetry, it is possible to define and compute through localization a supersymmetric sphere partition function $Z_{S^3}(m)$. The partition function depends on a choice of R-symmetry, i.e. a linear combination of the UV R-symmetry and of other Abelian flavor symmetries of the theory with coefficients $m$. 
If the infrared R-symmetry is not accidental, it should coincide with a choice of R-symmetry which minimizes $|Z_{S^3}(m)|^2$, i.e. maximizes $F(m) =-\log |Z_{S^3}(m)|$ \cite{Jafferis:2010un,Closset:2012vg}. 
This is compatible with the F-theorem: superpotential deformations of the theory which break some flavor symmetry do not change the value of the trial sphere partition function, 
but reduce the number of parameters available in the maximization. 

In this note we would like to discuss the conjecture that both results should
generalize to three-dimensional conformal boundary conditions for four-dimensional CFTs.
Given such a boundary condition, we can compute partition functions both on the four-sphere $S^4$ and on the hemisphere $HS^4$ (with boundary $S^3$), and define a boundary free energy $F_{\partial}$ as 
\begin{equation}
\frac{|Z_{HS^4}|^2}{Z_{S_4}} = e^{a_3 \frac{R^3}{\epsilon^3} + a_1 \frac{R}{\epsilon} - 2 F_{\partial}}
\end{equation}
The divergences in the partition function which have a four-dimensional origin will cancel out in the ratio. 
Although in the presence of boundaries in four dimensions one has extra logarithmic divergences proportional to powers of the extrinsic curvature (see \cite{Vassilevich:2003xt} for a review), the boundary of the hemisphere has no extrinsic curvature. The possibility of extra logarithmic divergences at the boundary is the reason we use the hemisphere rather than a ball, whose boundary would have extrinsic curvature. 

The remaining divergences should have a form similar to the divergences for a three-dimensional theory, and in particular 
the constant term should not suffer of logarithmic ambiguities, but only contain the power-law divergences subtracted explicitly in the above formula. 

For half-BPS boundary conditions in $\CN=2$ four-dimensional SCFTs, which preserve an $\CN=2$ three-dimensional super algebra, 
the hemisphere partition function is computable through localization \cite{Drukker:2010jp}. 
It will be functions of a choice of IR R-symmetry, a mixture of the Cartan 
sub algebra of the bulk $SU(2)_R$ symmetry and of boundary flavor symmetries.

We would like thus to conjecture that
\begin{itemize}
\item The boundary free energy $F_{\partial}$ decreases along boundary RG flows between conformal boundary conditions for the same CFT.
\item The exact infrared R-symmetry of a superconformal boundary condition maximizes $F_{\partial}(m)$.
\end{itemize}

These statements have an obvious extension to interfaces, simply by the doubling trick. For completeness, we can write an explicit formula for the boundary free energy of an interface between two theories $A$ and $B$, in terms of $S^4$ partition functions for either theory or in the presence of the interface:
\begin{equation}
\frac{|Z_{S^4}^{AB}|^2}{Z^A_{S_4}Z^B_{S_4}} = e^{a_3 \frac{R^3}{\epsilon^3} + a_1 \frac{R}{\epsilon} - 2 F_{\partial}}
\end{equation}

A similar boundary F-theorem has already been formulated in terms of entanglement entropy 
for a hemisphere centred on the boundary or interface in \cite{Jensen:2013lxa,Estes:2014hka} and tested in an holographic context. 
It should be straightforward to match the free-energy based definition of $F_{\partial}$ with the definition given in terms of entanglement entropy. 
It should also be possible to prove the boundary F-theorem along the same lines as in \cite{Casini:2012ei}.
In a similar fashion, the standard arguments given in support of F-maximization for $3d$ theories 
can be adapted to the context of a boundary or interface. 

Rather than pursuing a proof along these lines, in this note we will simply present a few very basic examples, 
which test the boundary F-theorem and boundary F-maximization principle for several weakly-coupled systems.  

\section{A perturbative proof}
It is straightforward to analyze the case of a BCFT perturbed by one or more slightly relevant boundary operators. 
As the two and three-point correlation functions of boundary operators have the same functional forms as in a 3d CFT, 
the calculation of the change in the hemisphere partition function is completely analogous to the calculation in sections 2 and 3 of 
\cite{Klebanov:2011gs} and the same monotonicity results apply here.  

\section{A free example}
The simplest example we can look at are Dirichlet and Neumann boundary conditions for a conformally coupled free scalar. 
The one-loop determinant in  
\begin{equation}
F=-\log Z = \frac{1}{2} \sum_n d_n \log \frac{\lambda_n}{R^2 \mu^2}
\end{equation}
with $d_n$ being the multiplicity of the eigenvalue $\lambda_n$ of the conformal Laplacian on the sphere, 
is usually regularized by using the function  
\begin{equation}
f(s) = \frac{1}{2} \sum_n \frac{d_n}{\lambda_n^s}
\end{equation}
as
\begin{equation}
F = \left[ - \partial_s f(s) - f(s) \log R^2 \mu^2 \right]|_{s=0}
\end{equation}

For a four-sphere, $\lambda_n = (n+1)(n+2)$ and $d_n = \frac{1}{6}(2n+3)(n+2)(n+1)$. 
It is convenient to do the computation with a slightly different regularization
\begin{equation}
\tilde f(s) = \frac{1}{2} \sum_n \frac{d_n}{(n+1)^s}+\frac{1}{2} \sum_n \frac{d_n}{(n+2)^s}
\end{equation}

We can compute 
\begin{align}
\sum_n \frac{d_n}{(n+a)^s} &= \frac{1}{3} \zeta(s - 3, a) - \frac{-3 + 2 a}{2} \zeta(s - 2, a) + \cr &+\frac{13 - 18 a + 6 a^2}{6} \zeta(s - 1, a) + \frac{6 - 13 a + 9 a^2 - 2 a^3}{6} \zeta(s, a)\end{align}
Replacing $f$ with $\tilde f$ in $F$ gives 
\begin{equation}
F_{S^4}= \frac{1}{90}\log R^2 \mu^2- \frac{1}{6} \zeta'(-1)- \frac{1}{3} \zeta'(-3)
\end{equation}

For an hemisphere with Dirichlet boundary conditions, we should only keep the spherical eigenfunctions which are odd under reflection across the equator. 
For Neumann, we should keep the even ones. We have $d^D_n =  \frac{1}{6}(n+2)(n+1)n$ and $d^N_n =  \frac{1}{6}(n+3)(n+2)(n+1)$.
We can compute 
\begin{align}
\sum_n \frac{d^D_n}{(n+a)^s} &= \frac{1}{6} \zeta(s - 3, a) - \frac{-1 + a}{2} \zeta(s - 2, a) +\cr &+\frac{(2 - 6 a + 3 a^2)}{6} \zeta(s - 1, a) + \frac{-2 a + 3 a^2 - a^3}{6} \zeta(s, a)\end{align}
and 
\begin{equation}
F^D_{HS^4}= \frac{1}{2} F_{S^4} -\frac{\zeta(3)}{16 \pi^2}
\end{equation}
i.e. 
\begin{equation}
F^D_{b}= -\frac{\zeta(3)}{16 \pi^2} \sim -0.00761211
\end{equation}

Clearly, for Neumann b.c we have the opposite value 
\begin{equation}
F^N_{b}= \frac{\zeta(3)}{16 \pi^2} \sim 0.00761211
\end{equation}

This is compatible with the conjectured $F_{\partial}$-theorem: Neumann b.c. flow to Dirichlet b.c. under 
deformation by a boundary operator $\Phi^2$ quadratic in the bulk scalar field $\Phi$. 
Indeed, $F^N_{\partial} > F^D_{\partial}$. 

On the other hand, if we take Dirichlet boundary conditions for $\Phi$, add an extra 
free three-dimensional scalar $\phi$ at the boundary, we obtain a boundary condition 
which can be deformed back to Neumann by adding a boundary coupling $\phi \partial_n \Phi$. 
As the free energy of a free 3d scalar is 
\begin{equation}
F_{3d}= \frac{1}{8}\log 2 - \frac{3\zeta(3)}{16 \pi^2} \sim 0.0638
\end{equation}
we see that $F^D_{\partial} +F_{3d} > F^N_{\partial}$, as it should.  

\section{Boundary conditions for free fields}
Consider a 3d CFT which contains some scalar operator $\CO$ of dimension $2-\eta$ for some small number $\eta$.
We can couple it to a four-dimensional free scalar $\Phi$ with Neumann b.c. by the linear boundary coupling 
\begin{equation}
S_{\partial} = g \int d^3x \Phi \CO
\end{equation}  
Roughly, the effect of this interaction is to deform the Neumann b.c. 
to
\begin{equation}
\partial_\perp \Phi|_\partial = g \CO
\end{equation}

For a conformally invariant boundary condition, the OPE of a free field $\Phi$ to the boundary is non-singular 
and the boundary values of $\Phi$ and $\partial_\perp \Phi$ are boundary operators of dimensions exactly $1$ and $2$ 
\footnote{This is a straightforward consequence of the boundary OPE expansion and the bulk equations of motion. A more detailed discussion can be found, say, in \cite{Dimofte:2012pd}}. 
Thus if the RG flow initiated by the deformation $S_{\partial}$ ends at some interacting conformal-invariant boundary condition,
the anomalous dimension of $\CO$ must shift to $2$. For small $\eta$, we can hope to find a perturbative fixed point with 
anomalous dimensions of boundary operators which differ at order $\eta$ from the UV dimensions.

There is a useful way to map this problem to a more standard three-dimensional problem. 
The three-dimensional effect of the bulk scalar field is essentially captured by a non-local bilinear coupling 
\begin{equation}
S_{\partial}^{\mathrm{eff}} = g^2 \int d^3x \int d^3y G_{\partial}(x,y) \CO(x) \CO(y)
\end{equation}
where 
\begin{equation}
G_{\partial}(x,y) \sim \frac{1}{(x-y)^2}
\end{equation}
is the restriction to the boundary of the propagator for $\Phi$ with Neumann b.c. 

We can mimic a similar interaction in a purely three-dimensional setup: couple the original 3d CFT 
to $N$ 3d free scalars $\phi_i$ through an interaction 
\begin{equation}
S_{3d} = \frac{g_{3d}}{\sqrt{N}} \int d^3 x \phi^2 \CO
\end{equation}
Indeed, in the large $N$ limit, the leading effective interaction induced by the 3d scalar fields is 
the same as $S_{\partial}^{\mathrm{eff}}$, with $g \sim g_{3d}$ up to a numerical factor.

At least at the level of perturbation theory, the decrease in $F_{\partial}$ due to turning on the interaction $S_{\partial}$ should be the same as 
the decrease of $F$ for this auxiliary theory at the leading order in $N$  due to turning on the interaction $S_{3d}$, and in particular it should be positive. 

Similar considerations apply for boundary conditions for free Abelian gauge fields. 
If we are given a 3d CFT with a $U(1)$ flavor symmetry, we can couple as boundary degrees of freedom for a 4d Abelian gauge field 
with Neumann boundary conditions. That modifies the Neumann boundary conditions to 
\begin{equation}
g^{-2} F_{i \perp} = J^{\mathrm{3d}}_i
\end{equation}
where $J^{\mathrm{3d}}_i$ is the $U(1)$ conserved current for the boundary theory and $g^2$ the coupling constant.
As the coupling is part of the bulk Lagrangian, it will not run. Typically, we will expect that for small $g^2$, the boundary condition will flow to a 
BCFT which is perturbatively close to the decoupled 3d CFT. 

Again, at a perturbative level the effect of the bulk gauge fields can be accounted for by an effective interaction 
\begin{equation}
S_{\partial}^{\mathrm{eff}} = g^2 \int d^3x \int d^3y G^{\partial}_{\mu \nu}(x,y) J^\mu(x) J^\nu(y)
\end{equation}
with
\begin{equation}
G^{\partial}_{\mu \nu}(x,y) \sim \frac{\eta_{\mu \nu}}{(x-y)^2}
\end{equation}
We can mimic such an interaction by replacing the bulk gauge field by some judicious choice of 3d fields: 
an auxiliary 3d $U(1)$ gauge field coupled to a large number $N$ of scalar fields of charge $q$ and to the original 3d CFT.
Integrating away the $N$ scalar fields produces an effective propagator for the 3d gauge field which mimics $G^{\partial}_{\mu \nu}(x,y)$.
Thus, the perturbative behaviour of $F_{\partial}$ is related to the large $N$ behaviour of $F$ in the auxiliary 3d theory and $F_{\partial}$ should decrease upon coupling 
the boundary degrees of freedom to the 4d gauge fields. 
We will revisit this construction in a supersymmetric setting.

\section{$\CN=2$ theories}
Next, we can look at $\CN=2$ supersymmetric theories, which are amenable of a localization analysis. In the process, we will also learn a few more useful facts about non-supersymmetric examples. 

\subsection{Free hypermultiplet}
First, we can consider a single bulk hypermultiplet. This theory has a $Sp(1)_f$ flavor symmetry, which together with the $SU(2)_R$ 
symmetry rotates the four real scalar fields. In order to define half-BPS boundary conditions, one can split the scalars into two complex fields $X$ and $Y$, 
with charge $1$ and $-1$ respectively under the $U(1)_f$ Cartan sub-algebra of the flavor symmetry and the same charge under the $U(1)_R$ 
Cartan sub-algebra of the R-symmetry preserved by the boundary condition. 

The restriction to a boundary or interface of $X$ and $Y$ is a 3d chiral operator, whose conformal dimension is fixed to $1$ by the bulk symmetries. 
Standard boundary conditions set to zero half of the fermions at the boundary and give either Dirichlet b.c. to $X$, Neumann b.c. to $Y$ or viceversa \cite{Drukker:2010jp}. 
We can denote these boundary conditions as ``$B_X$'' or ``$B_Y$'' depending on which field is set to zero at the boundary. 
If we have a free-hypermultiplet theory both on the positive and on the negative half-spaces, with $B_X$ boundary conditions on the negative side, 
$B_Y$ on the positive side, we can ``glue back'' the two halves by adding an interface superpotential 
\begin{equation}
W = X_{\partial^+} Y_{\partial^-}.
\end{equation}

Superpotential couplings do not affect localization computations. By the symmetries of the system, it follows that 
\begin{equation}
Z_{S_4}= Z_{HS^4}[B_Y] \bar Z_{HS^4}[B_X] = |Z_{HS^4}[B_X]|^2 
\end{equation}
i.e. $F_{\partial}[B_X] = 0$. 

This has the following implication, useful for non-supersymmetric computations: the contribution of the two real scalars with Dirichlet b.c. 
cancels out the contribution of the two real scalars with Neumann b.c. and thus the fermion contribution must be zero by itself. 
Thus if we have a bulk free fermion $\lambda_\alpha$ with a standard b.c. $B_\lambda: \mathrm{Re}\lambda_\alpha=0$, 
we must have $F_{\partial}[B_\lambda] = 0$.

We can obtain richer half-BPS boundary conditions by adding 3d degrees of freedom to a boundary with $B_Y$ b.c, and coupling them to the bulk hyper 
by an extra superpotential coupling 
\begin{equation}
W_{\partial} = X_{\partial} \CO.
\end{equation}
involving a 3d chiral operator $\CO$ with dimension smaller or equal to $1$. This setup can flow in the IR to a superconformal boundary condition. 

The main constraint which follows from $W_{\partial}$ is that the trial $U(1)$ R-charge of the operator $\CO$ must be such that its conformal dimension is fixed to $1$. 
Thus if $F(\Delta_a)$ is the $S^3$ partition function for the 3d degrees of freedom as a function of the trial $R$-charge assignemnts, we expect 
$F_{\partial}$ to be the maximum possible value of $F(\Delta_a)$ under the constraint $\Delta_\CO=1$. In particular, $F_{\partial}$ should be lower or equal than 
the value in the absence of the superpotential deformation $W_{\partial}$, which is the unconstrained maximum of $F(\Delta_a)$.

We can use the large $N$ argument in the previous section to convince ourselves that, at least perturbatively, this prescription for $F_{\partial}$ 
gives the correct answer. The effect of the bulk hypermultiplet on the boundary dynamics is captured by the boundary propagators for the bulk fields,
which are roughly the square of free 3d propagators. Thus the effect of $W_{\partial}$ should be similar to the large $N$ effect of a coupling to $N$ 3d chiral fields
 \begin{equation}
W_{3d} = \frac{1}{\sqrt{N}} \CO \phi^a \phi^a.
\end{equation}
This would give the same restriction on $\Delta_\CO$. 

\subsection{Free Abelian vectormultiplet}
The localization result for the hemisphere partition function with supersymmetric Dirichlet b.c. is 
\begin{equation}
Z_{HS^4}^D(\hat a) = e^{ - i \pi \tau \hat a^2}
\end{equation}
where $\hat a = i a + \delta$ is the combination of the vev $a$ fixed by the Dirichlet boundary conditions on the 
imaginary part of the scalar field in the vectormultiplet and a shift $\delta$ which accounts for the possibility 
that the $U(1)_g$ flavor symmetry at the boundary which arises from the bulk gauge symmetry may enter the trial R-symmetry. 
We set the overall normalization of the answer to $1$ because a constant normalization factor  would drop out of subsequent calculations. 

It is useful to  keep in mind the following facts:
\begin{itemize}
\item Two half-spaces with Dirichlet b.c. can be ``glued back'' by gauging in 3d the diagonal $U(1)_g$ flavor symmetry. 
\item Neumann b.c. can be obtained from Dirichlet b.c. by gauging the $U(1)_g$ boundary flavor symmetry. 
\item Dirichlet and Neumann b.c. are related by electric-magnetic duality.
\end{itemize}
In the presence of a theta angle, we refer to Neumann boundary conditions as 
\begin{equation}
\frac{2 \pi}{g^2} F_{3 i} = \frac{\theta}{4 \pi} \epsilon_{ijk} F^{jk} 
\end{equation}
Adding a Chern-Simons coupling at the boundary is equivalent to an integral shift in $\theta$. 

Localization tells us that gauging the 3d flavor symmetry coincides with a Fourier transform of the partition function,
\begin{equation}Z'(a') = \int da e^{ - 2 \pi i a a'} Z(a)
\end{equation}
where $a'$ is the mass parameter corresponding to the topological $U(1)$ symmetry. 
Thus we find for Neumann b.c. 
\begin{equation}
Z_{HS^4}^N(\hat a') =  \int da e^{ - 2 \pi i a a'+ i \pi \tau a^2} = \frac{1}{\sqrt{-i \tau}} e^{- \frac{i \pi}{\tau} (a')^2}
\end{equation}
and for the four-sphere
\begin{equation}
Z_{S^4} = \int da e^{ -2 \pi ( \mathrm{Im}\tau ) a^2} = \frac{1}{\sqrt{2 \mathrm{Im} \tau}} 
\end{equation}

The four-sphere partition function transforms as
\begin{equation}
Z_{S^4}(-\frac{1}{\tau}) = \tau^{1/2} \bar \tau^{1/2} Z_{S^4}(\tau)
\end{equation}
This anomalous transformation law is well known \cite{Witten:1995gf}.

The hemisphere partition functions transform as 
\begin{align}
Z_{HS^4}^N(a,-\frac{1}{\tau}) &= (- i \tau)^{1/2} Z_{HS^4}^D(a,\tau) \cr
Z_{HS^4}^D(a,-\frac{1}{\tau}) &= (- i \tau)^{1/2} Z_{HS^4}^N(a,\tau)
\end{align}
We expect the extra $(- i \tau)^{1/2}$ pre-factor to be universal in S-duality transformations of hemisphere partition functions. 
In particular, it cancels out of $F_{\partial}$, which transforms as a function on the space of gauge couplings: 
\begin{align}
F_{\partial}^N(a,-\frac{1}{\tau}) &= F_{\partial}^D(a,\tau) \cr
F_{\partial}^D(a,-\frac{1}{\tau}) &= F_{\partial}^N(a,\tau)
\end{align}

If we include $\CN=2$ 3d matter on the boundary, with $S^3$ partition function $Z_{3d}(\hat a, \Delta_i)$, the hemisphere partition function should be 
\begin{equation}
Z_{HS^4}(a',\Delta_i) =  \int da e^{ - 2 \pi i a a'+ i \pi \tau a^2} Z_{3d}(a, \Delta_i)
\end{equation}

Again, we can give an alternative interpretation of this formula, which relates $F_{\partial}$ maximization to $F$-maximization in a
related, purely three-dimensional setup. Imagine coupling the 3d degrees of freedom to a 3d Abelian Chern-Simons theory, 
with level $k$, which is also coupled to $N$ 3d scalar fields of charge $q$.
That would give, in the notations of \cite{Jafferis:2010un},
\begin{equation}
Z_{S^3}^{CS}(a',\Delta_i) =  \int da e^{ - 2 \pi i a a'+ i \pi k a^2 + N \ell(\frac{1}{2}+ i q a)} Z_{3d}(a, \Delta_i)
\end{equation}
Perturbatively, we can take a limit of large $N$ with $N q^2$ finite, and $N \ell(\frac{1}{2}+ i q a) \to N q^2 \frac{\pi^2 a^2}{2}$.
Thus the $N q^2$ coupling simulates the 4d gauge coupling, and the 3d CS coupling simulates the 4d $\theta$ angle. 

This has a simple physical meaning: the effect of the 4d gauge field on the boundary degrees of freedom is captured by the 
boundary propagator. In the presence of a $\theta$-term, the gauge theory boundary propagator is identical to 
the propagator for a 3d gauge theory with a Chern-Simons interaction $\theta/(2 \pi)$ and a non-local kinetic term of the same form as the one 
which would be produced by integrating out the $N$ 3d chiral fields above. This confirms again that, at least perturbatively, 
$F_b$ maximization is equivalent to $F$-maximization for an appropriate 3d field theory. 

\subsection{Non-Abelian bulk SCFTs}
The hemisphere partition function for a general non-Abelian $\CN=2$ SCFT with a UV Lagrangian description has the general form 
\begin{equation}
Z_{HS^4}(\Delta_i) \int d\nu_a Z_{4d}^N(a,m_c, \tau) Z_{3d}(a,m_c, \Delta_i)
\end{equation}
where we denoted as $m_c$ the values of the bulk mass parameters corresponding to a conformal coupling on the four-sphere, 
and with $\Delta_i$ the boundary flavor symmetries which are not fixed by the superpotential couplings to bulk hypermultiplets. 
The $Z_{4d}^N(a,m_c, \tau)$ part combines the tree-level, one-loop and instanton contributions to the partition function of the bulk theory. 
The $Z_{3d}(a,m_c, \Delta_i)$ part is the standard $S^3$ partition function for the boundary degrees of freedom. 

For a bulk $\CN=4$ gauge theory, much the same considerations apply as for a free Abelian gauge theory. The instanton contributions to the hemisphere 
partition functions cancel out if we set the adjoint mass to the conformal value. The one-loop factors reduce to the familiar Vandermonde-like determinant 
$\prod_\alpha \sinh^2 2 \pi (\alpha \cdot a)$ one encounters also in the calculation of $S^3$ partition functions. 

For general, non-Abelian $\CN=2$ gauge theories, the integrand is much richer, and includes bulk instanton corrections. We leave a full investigation 
of the $F^\partial$ maximization conjecture for these theories to later work.

\section*{Acknowledgements}
DG is grateful to Daniel Jafferis and Rob Myers for useful conversations and comments on the draft. 
The research of DG was supported by the Perimeter Institute for Theoretical Physics. Research at Perimeter Institute is supported by the Government of Canada through Industry Canada and by the Province of Ontario through the Ministry of Economic Development and Innovation. 

\bibliographystyle{JHEP}

\bibliography{boundaryF}

\providecommand{\href}[2]{#2}\begingroup\raggedright\begin{thebibliography}{10}

\bibitem{Myers:2010xs}
R.~C. Myers and A.~Sinha, {\it {Seeing a c-theorem with holography}},  {\em
  Phys.Rev.} {\bf D82} (2010) 046006,
  [\href{http://arxiv.org/abs/1006.1263}{{\tt arXiv:1006.1263}}].

\bibitem{Myers:2010tj}
R.~C. Myers and A.~Sinha, {\it {Holographic c-theorems in arbitrary
  dimensions}},  {\em JHEP} {\bf 1101} (2011) 125,
  [\href{http://arxiv.org/abs/1011.5819}{{\tt arXiv:1011.5819}}].

\bibitem{Jafferis:2011zi}
D.~L. Jafferis, I.~R. Klebanov, S.~S. Pufu, and B.~R. Safdi, {\it {Towards the
  F-Theorem: N=2 Field Theories on the Three-Sphere}},  {\em JHEP} {\bf 1106}
  (2011) 102, [\href{http://arxiv.org/abs/1103.1181}{{\tt arXiv:1103.1181}}].

\bibitem{Casini:2012ei}
H.~Casini and M.~Huerta, {\it {On the RG running of the entanglement entropy of
  a circle}},  {\em Phys.Rev.} {\bf D85} (2012) 125016,
  [\href{http://arxiv.org/abs/1202.5650}{{\tt arXiv:1202.5650}}].

\bibitem{Casini:2011kv}
H.~Casini, M.~Huerta, and R.~C. Myers, {\it {Towards a derivation of
  holographic entanglement entropy}},  {\em JHEP} {\bf 1105} (2011) 036,
  [\href{http://arxiv.org/abs/1102.0440}{{\tt arXiv:1102.0440}}].

\bibitem{Jafferis:2010un}
D.~L. Jafferis, {\it {The Exact Superconformal R-Symmetry Extremizes Z}},  {\em
  JHEP} {\bf 1205} (2012) 159, [\href{http://arxiv.org/abs/1012.3210}{{\tt
  arXiv:1012.3210}}].

\bibitem{Closset:2012vg}
C.~Closset, T.~T. Dumitrescu, G.~Festuccia, Z.~Komargodski, and N.~Seiberg,
  {\it {Contact Terms, Unitarity, and F-Maximization in Three-Dimensional
  Superconformal Theories}},  {\em JHEP} {\bf 1210} (2012) 053,
  [\href{http://arxiv.org/abs/1205.4142}{{\tt arXiv:1205.4142}}].

\bibitem{Vassilevich:2003xt}
D.~Vassilevich, {\it {Heat kernel expansion: User's manual}},  {\em Phys.Rept.}
  {\bf 388} (2003) 279--360, [\href{http://arxiv.org/abs/hep-th/0306138}{{\tt
  hep-th/0306138}}].

\bibitem{Jensen:2013lxa}
K.~Jensen and A.~O'Bannon, {\it {Holography, Entanglement Entropy, and
  Conformal Field Theories with Boundaries or Defects}},  {\em Phys.Rev.} {\bf
  D88} (2013) 106006, [\href{http://arxiv.org/abs/1309.4523}{{\tt
  arXiv:1309.4523}}].

\bibitem{Estes:2014hka}
J.~Estes, K.~Jensen, A.~O'Bannon, E.~Tsatis, and T.~Wrase, {\it {On Holographic
  Defect Entropy}},  \href{http://arxiv.org/abs/1403.6475}{{\tt
  arXiv:1403.6475}}.

\bibitem{Klebanov:2011gs}
I.~R. Klebanov, S.~S. Pufu, and B.~R. Safdi, {\it {F-Theorem without
  Supersymmetry}},  {\em JHEP} {\bf 1110} (2011) 038,
  [\href{http://arxiv.org/abs/1105.4598}{{\tt arXiv:1105.4598}}].

\bibitem{Dimofte:2012pd}
T.~Dimofte and D.~Gaiotto, {\it {An E7 Surprise}},  {\em JHEP} {\bf 1210}
  (2012) 129, [\href{http://arxiv.org/abs/1209.1404}{{\tt arXiv:1209.1404}}].

\bibitem{Drukker:2010jp}
N.~Drukker, D.~Gaiotto, and J.~Gomis, {\it {The Virtue of Defects in 4D Gauge
  Theories and 2D CFTs}},  {\em JHEP} {\bf 1106} (2011) 025,
  [\href{http://arxiv.org/abs/1003.1112}{{\tt arXiv:1003.1112}}].

\bibitem{Witten:1995gf}
E.~Witten, {\it {On S duality in Abelian gauge theory}},  {\em Selecta Math.}
  {\bf 1} (1995) 383, [\href{http://arxiv.org/abs/hep-th/9505186}{{\tt
  hep-th/9505186}}].

\end{thebibliography}\endgroup

\end{document}